# Synthetic aperture phase imaging of second harmonic generation field with computational adaptive optics

Jungho Moon[1,2,†], Sungsam Kang[1,2,†], Jin Hee Hong[1,2], Seokchan Yoon[1,2,3,*], and Wonshik Choi[1,2,*]

[1]*Center for Molecular Spectroscopy and Dynamics, Institute for Basic Science, Seoul 02841, Korea*
[2]*Department of Physics, Korea University, Seoul 02855, Korea*
[3]*School of Biomedical Convergence Engineering, Pusan National University, Yangsan 50612, Korea*
[†]*These authors contributed equally to this work*
*Correspondence addressed to sc.yoon@pusan.ac.kr and wonshik@korea.ac.kr*

**Abstract**
Second-harmonic generation (SHG) microscopy provides label-free imaging of biological tissues with unique contrast mechanisms, but its resolution is limited by the diffraction limit. Here, we present the first experimental demonstration of super-resolution quantitative phase imaging of the SHG field based on synthetic aperture Fourier holographic microscopy. We discuss the mathematical model of synthetic-aperture imaging of SHG fields, as well as the computational adaptive optics technique for correcting sample-induced aberration. We demonstrate proof-of-concept experiments where SHG targets are embedded within a thick scattering medium to validate the performance of the proposed imaging technique. It is shown to be able to overcome the conventional Abbe diffraction limit even in the complex aberrations and strong multiple scattering. We also demonstrate SHG-based super-resolution deep-tissue phase imaging of ex-vivo zebrafish muscle tissue.

**Introduction**
Over the past two decades, second harmonic generation (SHG) microscopy has emerged as a powerful tool for label-free, high-resolution imaging of biological tissues and materials, with unique contrast mechanisms that hold great potential for various applications in biomedical research and materials science. Unlike multi-photon excitation fluorescence (MPEF) microscopy (e.g., two-photon excitation[1] and three-photon excitation[2]) that requires exogenous fluorophores or labels, SHG microscopy provides inherent contrast based on the nonlinear optical properties of specimens, offering distinct advantages such as label-free imaging, reduced photobleaching, and unique sensitivity to endogenous molecular organization and non-centrosymmetric structures such as collagen fibers, microtubules, and muscular myosin.

A notable differentiation is that SHG is a coherent nonlinear scattering process that maintains long-range phase relations in second-harmonic fields, while MPEF relies on an incoherent process that involves noncoherent emission subsequent to multi-photon absorption. Due to the coherent nature of SHG, it is possible to obtain both amplitude-contrast and quantitative phase images of samples through interferometric measurements of SHG signals. The first interferometric measurement of SHG signal was conducted on nonlinear crystal nanoparticles[3]. Subsequently, SHG off-axis digital holographic microscopy (SHG-DHM) was utilized in experiments to distinguish between two types of nanoparticles [4] and to investigate second-harmonic signals generated at the glass/air interface [5]. After that, the application has been expanded to include holographic SHG quantitative phase imaging (QPI) of biological specimens, such as muscle structure[6], and collagen structure[7] in tissue. Unlike the fundamental phase measured by conventional DHM, which only provides information about the optical path length of an object, SHG phase provides additional structural information from second-order nonlinear susceptibility $\chi^{(2)}$.

Despite the narrowing of the excitation point spread function (PSF) due to the nonlinear process where the SHG signal scales as the square of the excitation intensity, the resolution of SHG microscopy is limited by the diffraction limit given by $\lambda_{SHG}/2NA$. Here $\lambda_{SHG}$ is the second-harmonic wavelength and NA is the numerical aperture of detection. Most of conventional SHG-DHM techniques mentioned above are based upon single-shot, wide-field illumination, and thus the lateral resolution is still diffraction-limited. To circumvent this limitation, synthetic aperture Fourier holographic microscopy technique[8] that utilizes the SHG field has been suggested[9, 10]. Synthetic aperture Fourier holographic microscopy is also referred to as coherent structured illumination microscopy (SIM) achieve super-resolution QPI. Unlike conventional SIM for incoherent imaging, which employs spatially intensity-modulated patterns projected onto the sample, coherent SIM utilizes a series of phase-modulated illumination fields, which can be simply generated by tilting the illumination angle. An interferometric measurement is taken on the scattered SHG field for each illumination angle to determine the complex (amplitude and phase) image of the SHG field. These complex SHG images are then aperture-synthesized in the spatial-frequency domain to retrieve the object spectrum with an increased bandwidth, up to twice that of the detection optical system. Despite advancements in holographic imaging techniques, there has been no experimental demonstration of synthetic aperture phase imaging of the SHG field to date. The low efficiency of harmonic conversion poses technical difficulties, requiring the use of a high-power pulsed laser for wide-field illumination and a high-sensitivity detection system. Moreover, imaging deep inside biological tissue can lead to strong multiple-scattering noise and wavefront distortion due to sample-induced aberrations, which can easily overwhelm the image signal. These practical issues pose significant challenges for proper image reconstruction, rendering super-resolution imaging virtually unachievable. There also have not been much researches on the physical model and the principle of synthetic aperture to SHG imaging.

In this paper we present the first experimental demonstration, to the best of our knowledge, of super-resolution QPI of SHG field based on synthetic aperture Fourier holographic microscopy. We first describe the working principle of our newly-developed synthetic-aperture SHG (SA-SHG) microscopy, followed by a discussion on the mathematical model of the synthetic-aperture image formation. In particular, we address the challenges posed by a realistic imaging scenario where a thin SHG target of interest is embedded within a thick scattering medium, resulting in strong aberrations and multiple scattering. Next, we present a proof-of-concept experiment performed using a thin target structure made by WSe2 transition metal dichalcogenide (TMD) monolayers [11, 12]. Furthermore, we showcase the performance of our SA-SHG microscopy with an artificial phantom target in the presence of strong multiple scattering and complex aberrations. The SA-SHG imaging is based on coherent summation of complex SHG spectra, which inherently provides efficient suppression of multiple-scattering noise. Also, we apply a computational aberration correction method [13] to restore aberration-free super-resolution images. Finally, we demonstrate the SA-SHG imaging of ex-vivo zebrafish muscle.

**Results**
**Experimental schematic and working principle of the SA-SHG**
We developed a wide-field off-axis interferometric SHG microscopy as shown in Fig. 1a. In the setup, a regenerative amplifier laser (Coherent Libra, center wavelength 800 nm, pulse width 91 fs, repetition rate 10 kHz) is used to excite SHG signal of the sample over a wide range of the field-of-view. The typical illumination beam power entering the sample was 2.3 mW with a pulse energy of 0.09 nJ/µm². The angle of the incident plane wave was adjusted with a spatial light modulator (SLM, Hamamatsu, X13138-02). The SHG signal generated from the sample was delivered to a camera (PCO.edge4.2LT) after passing through a bandpass filter (center wavelength 400 nm) to filter out the excitation beam wavelength. For interferometric detection, the reference wave beam is combined with the sample SHG signal with an off-axis configuration after passing a beta barium borate (BBO, EKSMA optics, 0.5 mm thickness, type 1 phase matching condition) crystal. A grating was used in the reference arm to generate a first-order diffracted beam for single-shot off-axis measurement. This setup allows us to obtain the complex electric field of the sample SHG signal at a desired illumination angle in a single shot.

Let's consider a thick scattering medium and a thin object plane of interest within the medium as shown in Fig. 1b. If we send the initial field $E_0$ to the medium, the incident field $E_{\text{inc}}$ on the object plane can be described by the superposition of ballistic wave $E_B$ and the multiple scattering $E_M$ as $E_{\text{inc}}(\mathbf{r}, \omega) = E_B(\mathbf{r}, \omega) + E_M(\mathbf{r}, \omega)$, where $\omega$ is angular frequency of the incident field. The ballistic field $E_B$ is then given by the initial field with a ballistic attenuation coefficient $\gamma_1$ as $E_B = \gamma_1 E_0$. Then the SHG field $E_{\text{SHG,obj}}$ at the object plane can be described by the 2$^{\text{nd}}$ order nonlinear susceptibility $\chi^{(2)}(\mathbf{r})$ as,

$$E_{\text{SHG,obj}}(\mathbf{r}, 2\omega) = \chi^{(2)}(\mathbf{r}) E_{\text{inc}}^2(\mathbf{r}, \omega). \quad (1)$$

The detected signal at the camera plane is the superposition of the multiple-scattered SHG field $E_{\text{SHG,M}}$ with the ballistic SHG field $E_{\text{SHG,B}}$ that does not experience multiple scattering. By introducing another ballistic attenuation coefficient $\gamma_2$ in the transmission path, we can describe the SHG field as,

$$E_{\text{SHG}}(\mathbf{r}, 2\omega) = \gamma_1^2 \gamma_2 \chi^{(2)}(\mathbf{r}) E_0^2(\mathbf{r}, \omega) + E_{\text{SHG,M}}(\mathbf{r}, 2\omega). \qquad (2)$$

Without absorption, the attenuation coefficients $\gamma_1$ and $\gamma_2$ are a function of medium thickness as, $\gamma_1 = e^{-L_1/l_s(\omega)}$ and $\gamma_2 = e^{-L_2/l_s(2\omega)}$, respectively, where $L_1$ and $L_2$ are medium thickness before and after the object plane, respectively, and $l_s$ is the wavelength-dependent scattering mean free path. In Eq. (2), $E_{\text{SHG,M}}$ includes the SHG field generated by the multiple scattering components of $E_{\text{inc}}$ and the multiple scattering components occurring after SHG generation. Therefore, Eq. (2) accounts for both the multiple scattering occurring at the fundamental frequency $\omega$ and the SHG frequency $2\omega$.

We can obtain the spectrum of the SHG field by the Fourier transform of Eq. (2) as,

$$\tilde{E}_{\text{SHG}}(\mathbf{k}, 2\omega) = \gamma_1^2 \gamma_2 \tilde{\chi}^{(2)}(\mathbf{k}) \circledast_\mathbf{k} [\tilde{E}_0(\mathbf{k}, \omega) \circledast_\mathbf{k} \tilde{E}_0(\mathbf{k}, \omega)] + \tilde{E}_{\text{SHG,M}}(\mathbf{k}, 2\omega) \qquad (3)$$

where $\tilde{E}_{\text{SHG}}$, $\tilde{E}_0$ and $\tilde{E}_{\text{SHG,M}}$ are the Fourier transform of $E_{\text{SHG}}$, $E_{\text{in}}$ and $E_{\text{SHG,M}}$, respectively, and $\circledast_\mathbf{k}$ stands for the convolution in the spatial frequency domain. Under a plane wave illumination with a specific incident wavevector $\mathbf{k}_{\text{in}}$, the initial field can be written as, $\tilde{E}_0(\mathbf{k}, \omega) = P_{\text{in}}(\mathbf{k})\delta(\mathbf{k} - \mathbf{k}_{\text{in}})$ with input pupil function $P_{\text{in}}(\mathbf{k})$, and Eq. (3) becomes,

$$\tilde{E}_{\text{SHG}}(\mathbf{k}, 2\omega) = \gamma_1^2 \gamma_2 P_{\text{out}}(\mathbf{k})\tilde{\chi}^{(2)}(\mathbf{k} - 2\mathbf{k}_{\text{in}})P_{\text{in}}^2(\mathbf{k}_{\text{in}}) + \tilde{E}_{\text{SHG,M}}(\mathbf{k}, 2\omega) \qquad (4)$$

where $P_{\text{out}}(\mathbf{k})$ is the output pupil function. Ideally, $P_{\text{in}}(\mathbf{k})$ and $P_{\text{out}}(\mathbf{k})$ are circle functions that have unit values when $|\mathbf{k}| < k_0 NA_{\text{in}}$, and $|\mathbf{k}| < 2k_0 NA_{\text{out}}$, respectively. In the expression, $NA_{\text{in}}$ and $NA_{\text{out}}$ are the numerical aperture of the illumination and detection objective lenses, respectively, and $k_0$ is the vacuum wavenumber of the fundamental wave.

Finally, the nonlinear susceptibility $\tilde{\chi}^{(2)}$ can be factorized by mapping the measured SHG field after the lateral shift of $-2\mathbf{k}_{\text{in}}$ as,

$$\tilde{E}_{\text{SHG}}(\mathbf{k} + 2\mathbf{k}_{\text{in}}) = \gamma_1^2 \gamma_2 P_{\text{out}}(\mathbf{k} + 2\mathbf{k}_{\text{in}})\tilde{\chi}^{(2)}(\mathbf{k})P_{\text{in}}^2(\mathbf{k}_{\text{in}}) + \tilde{E}_{\text{SHG,M}}(\mathbf{k} + 2\mathbf{k}_{\text{in}}). \qquad (5)$$

In general, the intensity of $\tilde{E}_{\text{SHG,M}}$ is much stronger than that of the ballistic SHG field in a thick scattering medium, and the reconstruction process in Eq. (5) does not work properly. Therefore, we need to suppress the contribution of SHG multiple scattering with the additional measurement at different illumination angles. Comparing the SHG field at different illumination angles, the ballistic SHG fields have a certain correlation by the common $\tilde{\chi}^{(2)}$ distribution, whereas the SHG multiple scattering is completely uncorrelated. This means that we can exploit the coherent accumulation of ballistic SHG signals while suppressing the contribution of the multiple scattering[14]. If we define the output SHG field under $i^{\text{th}}$ illumination angle as $\tilde{E}_{\text{SHG}}(\mathbf{k}; \mathbf{k}_i)$ and perform $N$ number of measurements at different illumination angle, then the coherent accumulation process of Eq. (5) can be written as,

$$\sum_{i=1}^{N} \tilde{E}_{\text{SHG}}(\mathbf{k} + 2\mathbf{k}_i; \mathbf{k}_i) = \gamma_1^2 \gamma_2 N \cdot H(\mathbf{k})\tilde{\chi}^{(2)}(\mathbf{k}) + \sum_{i=1}^{N} \tilde{E}_{\text{SHG,M}}(\mathbf{k} + 2\mathbf{k}_i; \mathbf{k}_i). \qquad (6)$$

In the above equation, we defined $\sum_{i=1}^{N} P_{\text{out}}(\mathbf{k} + 2\mathbf{k}_i)P_{\text{in}}^2(\mathbf{k}_i) \equiv N \cdot H(\mathbf{k})$, with $H(\mathbf{k})$, the modulation transfer function of the system. The amplitude of the first term of Eq. (6) increases with the number of measurements $N$. On the contrary, the fluctuation of the multiple scattering components in the last term of Eq. (6) grows with $\sqrt{N}$ by the stochastic nature of random speckle field [15]. Therefore, the SNR of the $\tilde{\chi}^{(2)}$ reconstruction increases by $\sqrt{N}$, while the SNR of the SHG image intensity increases by $N$. It is noteworthy that this process is identical to that of a linear synthetic aperture microscope, except that the amount of spatial frequency shift is doubled for SHG due to the nonlinear process [16].

Figure 1c shows the coherent summation process in Eq. (6) for three specific incidence angles $(\mathbf{k}_1, \mathbf{k}_2, \mathbf{k}_3)$. First, each measured SHG spectrum $\tilde{E}_{\text{SHG}}(\mathbf{k}; \mathbf{k}_i)$ is shifted in the spatial frequency domain with an amount of $-2\mathbf{k}_i$. In the figure, each shaded circle corresponds to a shifted output pupil function $P_{\text{out}}(\mathbf{k} + 2\mathbf{k}_i)$ whose radius is given by $2k_0 NA_{\text{out}}$. Then, the summation of the three shifted SHG spectra are normalized by $H(\mathbf{k}) = \sum_{i=1}^{N} P_{\text{out}}(\mathbf{k} + 2\mathbf{k}_i) P_{\text{in}}^2(\mathbf{k}_i)/N$ to reconstruct the $\tilde{\chi}^{(2)}$ distribution in the spatial frequency domain. Finally, by inverse Fourier transform of $\tilde{\chi}^{(2)}$, we can reconstruct the image of $\chi^{(2)}$ distribution in position domain.

So far, we have discussed how to suppress the multiple scattering in SA-SHG microscopy. In the process, we have dealt with the input and output pupil function as simple circle functions. However, this is not true in the presence of optical aberration induced by the sample itself. In general, optical aberrations are described by a complex pupil function whose phase is modulated [17]. Then the input and output pupil functions are expressed as $P_{\text{in}}(\mathbf{k}) = \text{circ}[\mathbf{k}/(k_0 NA_{\text{in}})] \cdot e^{i\phi_{\text{in}}(\mathbf{k})}$ and $P_{\text{out}}(\mathbf{k}) =$

$\text{circ}[\mathbf{k}/(2k_0 NA_{\text{out}})] \cdot e^{i\phi_{\text{out}}(\mathbf{k})}$ for input and output aberrations $\phi_{\text{in}}$ and $\phi_{\text{out}}$, respectively, with $\text{circ}(\mathbf{k}) = 1$ for $|\mathbf{k}| \leq 1$. Without correcting for these aberrations, the summation in $H(\mathbf{k})$ cannot be accumulated well due to the phase shifts. In a linear imaging system, such an aberration problem can be solved by the algorithm termed closed loop accumulation of single scattering (CLASS). The functional form of Eq. (4) is identical to the formalism of the CLASS algorithm by letting $\mathbf{k}'_{\text{in}} \equiv 2\mathbf{k}_{\text{in}}$. Therefore, we can find the map of $\phi_{\text{in}}$ and $\phi_{\text{out}}$ by applying the CLASS algorithm to Eq. (4). Then, we can correct the aberrations by multiplying the complex conjugate of the acquired aberration maps in the left-hand-side of Eq. (4). After following the same processes in Eq. (5)-(6), we can finally obtain the aberration-corrected SA-SHG images.

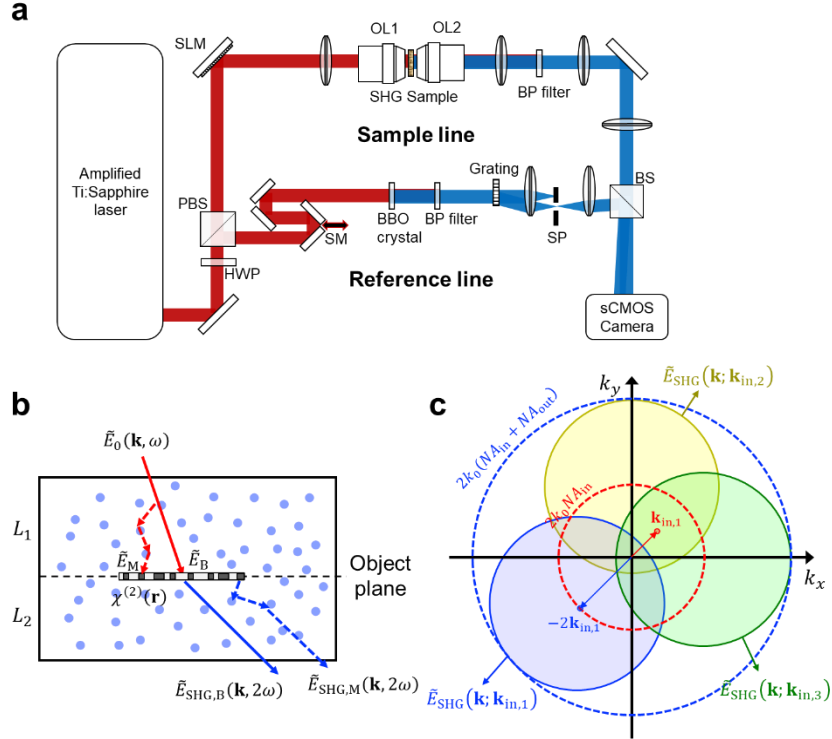

**Figure 1. Experimental setup and working principle of SA-SHG microscopy. a,** Schematic of SA-SHG microscopy. PBS: polarizing beam splitter, BS: beam splitter, SLM: spatial light modulator, HWP: half-wave plate, BP: band pass filter, OL1 and OL2: objective lenses, SM: scanning mirror, SP: spatial pinhole, and BBO crystal: Beta Barium Borate crystal. **b,** Schematic of SHG generation in scattering medium. **c,** Working principle of SA-SHG microscopy. Individual shaded circle corresponds to the spectrum of SHG field under incident plane wave with wavevector $(\mathbf{k}_{\text{in},1}, \mathbf{k}_{\text{in},2}, \mathbf{k}_{\text{in},3})$, respectively. Red dashed circle shows the maximum spectral shift $2k_0 NA_{\text{in}}$ in the synthetic aperture reconstruction of SA-SHG microscopy. Blue dashed circle shows the spectral coverage of $\tilde{\chi}^{(2)}(\mathbf{k})$.

**Experimental demonstration of the SA-SHG**

We experimentally demonstrated the SA-SHG microscopy with a thin target structure made by WSe2 transition metal dichalcogenide (TMD) monolayers (Ref). Lack of inversion symmetry in WSe2 allows strong SHG signal. We first measured the raw images of the SHG field with 300 ms exposure time under wide-field plane wave illumination. Total $N = 45$ incidence angles were scanned by the SLM. Next, we recovered the amplitude and the phase map of the SHG field by the Hilbert transform of the raw interferograms. Figure 2a-c shows the amplitude, phase, and the spectrum of the SHG field, respectively, at four representative angles of incidence. Then we combined the measured spectra after shifting each spectrum by $-2\mathbf{k}_{\text{in}}$ as shown in Fig. 2d. Without aberration, the sum of the shifted spectra in Eq. (6) is proportional to $H(\mathbf{k})\tilde{\chi}^{(2)}(\mathbf{k})$ with the modulation transfer function $H(\mathbf{k}) = \sum_{i=1}^{N} P_{\text{out}}(\mathbf{k} + 2\mathbf{k}_i)/N$. The shifted pupil function $P_{\text{out}}(\mathbf{k} + 2\mathbf{k}_i)$ is indicated by dashed circles in Fig. 2d. After the summation of Eq. (6), we could reconstruct $\chi^{(2)}(\mathbf{r})$ in position domain by taking the inverse Fourier transform after normalizing the $H(\mathbf{k})$. Figure 2e-f shows the amplitude and the phase map of the final SHG image of the sample under $N = 45$ incidence angles of illumination. The SA-SHG image shows much better image contrast compared to the single-shot images in Fig. 2a.

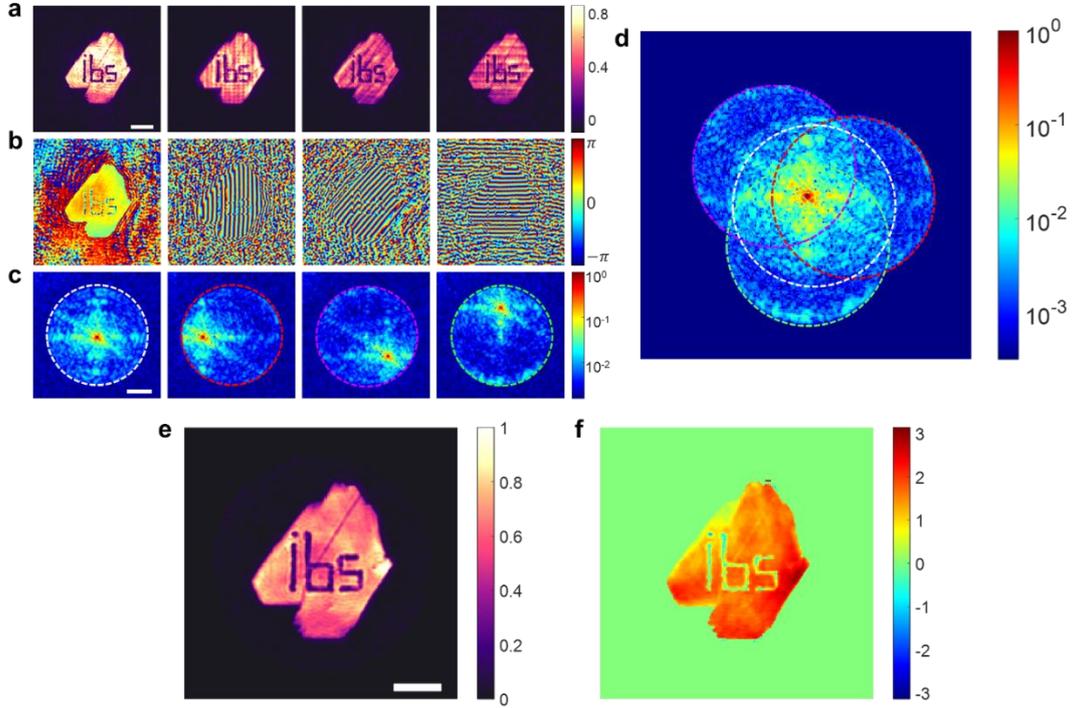

**Figure 2. Experimental demonstration of SA-SHG microscopy. a-c**, Amplitude, phase, and spectrum of the SHG field at four representative angles of incidence. Amplitude maps in **a** are normalized by the peak amplitude of the first image. Scale bar in **a**: 5 µm. Scale bar in **c**: $k_0 NA_{out}$ with $NA_{out} = 0.65$, and $k_0$ is the wavenumber of the fundamental wave. **d**, Reconstruction of $\tilde{\chi}^{(2)}(\mathbf{k})$ in spatial frequency domain. Dashed circles indicate the mapping position of SHG fields in **a**. The color scale in **d** is normalized by the peak amplitude and then displayed on a logarithmic scale. **e**, Amplitude map of the SA-SHG image reconstructed by $N = 45$ illumination angles. Color scale: normalized by the peak amplitude. Scale bar: 5 µm. **f**, Phase map of the SA-SHG image in **e**. Background phase outside the target structure is set to zero by applying an amplitude threshold level of 0.05. Color scale: phase in radian.

**Experimental demonstration of aberration correction in the SA-SHG microscopy**

In the previous section, we present the reconstruction process of the SA-SHG microscopy. Next, we demonstrate the performance of SA-SHG microscopy in the presence of multiple scattering and aberrations. As depicted in Fig. 3a, the same target structure used in Fig. 2 was embedded in a thick scattering medium. The scattering medium was made by ZnO and BaTiO3 nanoparticles mixed with PDMS. It has a thickness of 40 um with a scattering mean free path of 9 um for the 800 nm. By using batio3 nonlinear nanoparticles, not only elastic scattering but also SHG multiple scattering occurs. The sample is then sandwiched between two aberration layers made by lens cleaning gel (ESD-DF solution, First Contact Polymer Inc.). The random surface profile can result in complex aberrations even after the lens cleaning solution has dried and cured.

We measured the SHG field of the sample using $N = 300$ incidence angles of illumination. Figure 2b presents the amplitude of the SHG field under normal plane wave illumination. Due to the significant multiple scattering and aberrations, we could not identify any target structures. Nonetheless, by reconstructing the SA-SHG image from $N = 300$ images (Fig. 3c), we observed improved contrast as random multiple scattering was suppressed while coherent summation process in Eq. (6). However, the resulting SA-SHG image still lacks detailed structures because of the distortion caused by the aberrations. To correct these aberrations, we applied the CLASS algorithm to the measured SHG fields. After finding the input and output aberration maps $\phi_{\text{in}}$ and $\phi_{\text{out}}$, we multiplied the complex conjugate phase map to the initial SHG spectra. Then, we reconstructed the SA-SHG image again with the aberration corrected SHG fields. Figure 3d shows the resulting image. It shows fine details of the target structures with increased contrast. We also present the output aberration map $\phi_{\text{out}}(\mathbf{k})$ in Fig. 3e. Complicated phase distortion by the aberration is well resolved in the figure.

To quantitatively investigate the performance of the aberration correction, we analyzed the intensity enhancement of the coherent summation before and after the aberration correction. Without aberration and multiple scattering, the total intensity $I_{\text{total}} = \int \left| \sum_{i=1}^{N} \tilde{E}_{\text{SHG}}(\mathbf{k}+2\mathbf{k}_i;\mathbf{k}_i) \right|^2 d^2\mathbf{k}$ in Eq. (6) should be proportional to $N^2$. On the contrary, in case the multiple scattering in the last term of Eq. (6) is dominant, or the aberration is so severe that the MTF $H(\mathbf{k})$ become random, $I_{\text{total}}$ will be proportional to $N$ due to the nature of speckle field[15]. In other words, in the moderate scattering and aberration regime, $I_{\text{total}}$ will be proportional to $N^\alpha$ with $1 \leq \alpha \leq 2$. In Fig. 3f, we present the behavior of $I_{\text{total}}$ before and after aberration correction as a function of $N$. Before correction, $I_{\text{total}}$ was fitted with $\alpha = 1.612 \pm 0.0065$ which means that the scattering and aberration disturbs the coherent summation in Eq. (6). On the other hand, $\alpha = 1.989 \pm 0.0033$ after correction implies that the aberration is properly corrected.

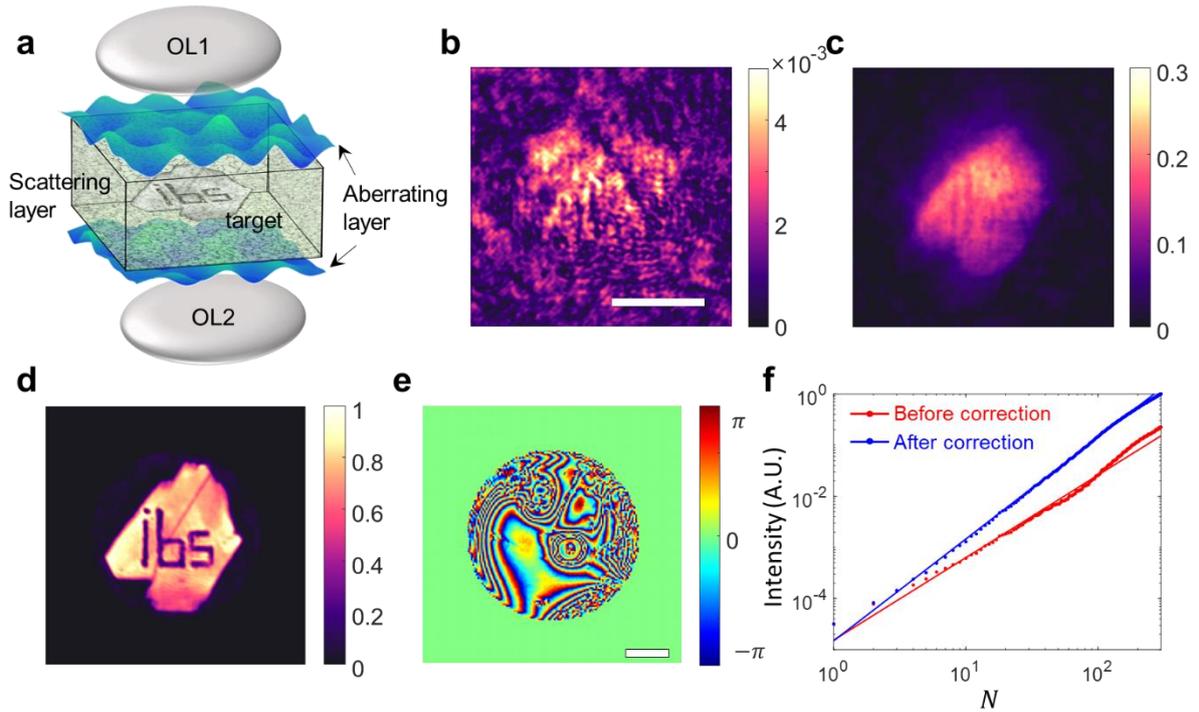

**Figure 3. SA-SHG imaging of a target in the presence of scattering and aberration. a**, Schematic of sample configuration. **b**, SHG field amplitude under a plane wave illumination. Scale bar: 10 µm. **c**, SA-SHG image under 300 illumination angles without aberration correction. **d**, SA-SHG image after aberration correction. Color scale of **b-d** are normalized by the peak amplitude of **d**. **e**, Quantified aberration map. Color scale: phase in radian. Scale bar: $k_0 NA_{\text{out}}$ with $NA_{\text{out}} = 0.65$, and $k_0$ is the wavenumber of the fundamental wave. **f**, Total intensity of SA-SHG image as a function of $N$ before and after aberration correction.

**Experimental demonstration of the SA-SHG microscopy with biological sample**
We have presented the working principle and performance of the SA-SHG microscope using a patterned target object and an artificial scattering and aberrating medium. In this section, we demonstrate the SA-

SHG microscopy on real biological samples. For the end, we prepared 9 dpf (days post fertilization) zebrafish fixed with paraformaldehyde in PBS.

We measured the SHG field of the sample with $N = 1,000$ incidence angles. Due to the severe multiple scattering and low SHG signal level, we require more images compared to the patterned target in Fig. 2-3. As shown in Fig. 4a, a single-shot SHG image does not show any structural information of the target. Meanwhile, the SA-SHG image reconstructed by $N = 1,000$ images shown in Fig. 4b show fine details of muscle tissues with much increased image contrast. However, it still shows only part of the tissue in detail due to the aberration caused by the body shape of the zebrafish. We found that the aberrations were non-uniform throughout the ROI due to the inhomogeneous body curvature of the zebrafish. To address this issue, we divided the entire ROI into 3×3 patches, and then applied the CLASS algorithm to each patch region. The result is shown in Fig. 4c-d. We constructed the aberration corrected SA-SHG images for each patch, and then combined the full ROI image. As shown in the figure, fine details of the muscle tissues are clearly visible across the whole ROI. We also present the locally varying aberration maps in Fig. 4d.

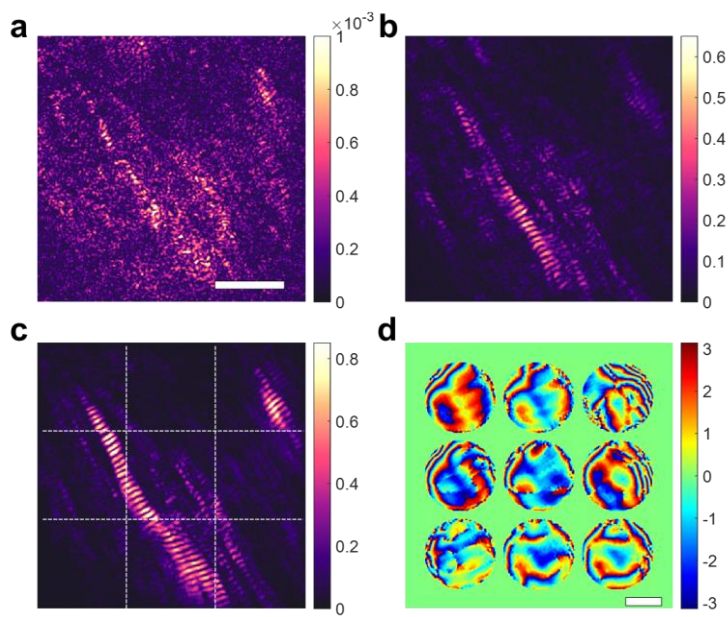

**Figure 4. SA-SHG imaging of a zebra fish muscle. a**, Amplitude of a wide-field SHG field image under normal plane wave illumination. Scale bar: 10 µm. **b**, Amplitude of SA-SHG image reconstructed from $N = 1,000$ illumination angles without aberration correction. **c**, Amplitude of SA-SHG image after aberration correction by $3 \times 3$ patches. White dashed line indicates the boundaries of each patch region. Color scale of **a**-**c** are normalized by the peak amplitude of **c**. **d**, Quantified aberration map. Color scale: phase in radian. Scale bar: $2k_0 NA_{\text{out}}$ with $NA_{\text{out}} = 0.65$, and $k_0$ is the wavenumber of the fundamental wave.

**Discussion**

SA-SHG imaging based on the coherent aperture image synthesis in the spatial-frequency domain, providing increased spatial-frequency bandwidth and efficient noise suppression. After aberration correction, we observed a linear increase in the total intensity of the synthesized SHG image with the number of illumination angles. However, the degree of SNR enhancement varied across spatial frequencies due to variations in the number of measurements taken for each spatial frequency. The number of measurements decreases linearly with increasing spatial frequency and eventually reaches one at the highest frequency. Consequently, the coherently synthesized object's spectrum exhibits relatively lower SNR at higher spatial frequencies, and the effective spatial-frequency bandwidth is determined by the maximum frequency at which SNR > 1. Thus, SNR becomes the predominant limiting factor for spatial resolution in the presence of strong background noise.

In this study, we assumed that the fundamental field had linear polarization, and the SHG field was detected with the same polarization angle. Consequently, the fundamental field and the second-order nonlinear susceptibility are regarded as scalars in Eq. (1). In general, due to the tensor nature of the second-order nonlinear susceptibility, the SHG signal is dependent on the polarization angle of the incident field relative to the spatial geometry of the sample of interest. The polarization-dependent nature of SHG signals has been extensively studied as a valuable modality for obtaining additional quantitative information about the geometric features of samples, such as the orientation of collagen fibers [18]. The SA-SHG microscopy can be readily expanded to polarization-resolved SHG (PSHG) imaging [19, 20] even with subcellular resolution. However, as the incident angle of fundamental field increases, the polarization component along the optical axis becomes significant and cannot be disregarded, resulting in incident angle-dependent SHG signals. Further studies are needed to investigate the effect of PSHG signals on synthetic-aperture image reconstruction in high-resolution SA-SHG imaging with a high illumination NA.

**Conclusion**

In conclusion, we have introduced SA-SHG microscopy, a novel super-resolution QPI technique that utilizes synthetic aperture Fourier holographic imaging of SHG fields. Our approach incorporates a novel computational adaptive optics strategy that corrects sample-induced aberrations through post-data processing. To confirm the validity of our proposed technique, we conducted validation experiments using SHG targets embedded in a highly scattering medium. Furthermore, we applied SA-SHG microscopy to observe zebrafish muscle tissue to demonstrate its capability for deep-tissue imaging in biological samples. We anticipate that SA-SHG microscopy will be a very promising technique for facilitating label-free, deep-tissue biological imaging and diagnostics.